\documentclass[aps,prl,superscriptaddress,showpacs,twocolumn,floatfix]{revtex4-1}
\usepackage{epsfig}
\usepackage{color}

\usepackage{scalerel}
\usepackage{tikz}
\usepackage{amsmath}
\usepackage{amssymb}
\usepackage{placeins}

\definecolor{red}{rgb}{1,0,0}
\definecolor{blue}{rgb}{0,0,1}
\definecolor{black}{rgb}{0,0,0}

\def\OO{{\cal O}}

\newcommand{\eq}[1]{\begin{align}#1\end{align}}

\newlength{\arrow}
\begin{document}

\title{Comment on: `What Determines the Static Force Chains in Stressed Granular Media?'}
\author{E. DeGiuli}
\affiliation{Institute of Theoretical Physics, Ecole Polytechnique F\'ed\'erale de Lausanne (EPFL), CH-1015 Lausanne, Switzerland}
\author{J. N. McElwaine}
\affiliation{Department of Earth Sciences, Durham University, Science Labs, Durham, DH1 3LE, U.K.}

\begin{abstract}
\end{abstract}
\maketitle


Generic packings of frictional disks or spheres are hyperstatic: there are more degrees of freedom in the contact forces than are fixed by the conditions of mechanical equilibrium. This implies that the history of packing preparation affects the internal structure and elastic response of frictional granular materials \cite{Agnolin07a,Atman13}. Such studies imply that a measure of packing fabric is necessary to model the solid behavior of granular materials, but the form of relevant internal variables is debated \cite{Edwards01,Ball02,DeGiuli14a}. In a recent Letter \cite{Gendelman16} , Gendelman, Pollack, Procaccia, Sengupta, and Zylberg (GPPSZ) claim to solve a strong version of this problem, presenting an equation from which the forces can be determined, given the positions of all particles and their radii. GPPSZ emphasize that their result (i) does not require the transverse force law, and (ii)  does not involve the distances between particle centers, since these cannot be precisely determined in experiments.  While their analysis, and claim (i), are correct, we show here that claim (ii) is false; in fact, the solution proposed by GPPSZ requires that particle radii and positions are known to the precision of the deformations at contacts, a feat impossible in experiments.

This result can simply be established by a scaling analysis of the equations in Ref.\onlinecite{Gendelman16}. We take units with the mean grain diameter $\langle \sigma \rangle = 1$, and rescale applied forces and torques by the pressure, $p$, which must also be the scale of the contact forces $|f\rangle$ to be determined. Then the main equation of Ref.~\onlinecite{Gendelman16} takes the form
\eq{ \label{1}
G | f \rangle = p \; \scaleleftright[3ex]{\Biggl |} { \; \begin{matrix} - |F^{ext} \rangle \\ \big(\frac{\kappa}{p}\big) Q | \sigma \rangle \end{matrix} \; } {\Biggr\rangle},
}
where $|F^{ext}\rangle$ is a vector of rescaled external forces and torques, $|\sigma \rangle$ is a vector of geometrical quantities linearly related to grain radii, $Q$ and $G$ are matrices involve $\OO(1)$ geometrical quantities, and $\kappa \gg p$ is the grain stiffness. The term involving $Q|\sigma \rangle$ contains the nontrivial geometrical constraints, one for each loop in the packing. For the linear elastic forces considered by GPPSZ, the quantity $\Delta \equiv p/\kappa$ is the typical magnitude of grain deformations; for a typical experiment, $\Delta \lesssim 10^{-5}$ \footnote{For Hertzian contacts, $\Delta=(p/E)^{2/3}$, where $E$ is the grain Young modulus. Within a 20cm deep sandpile, $\Delta~\approx~10^{-5}$}. Thus from \eqref{1} one would naively expect that either (i) $|f \rangle \sim p/\Delta$, which is far too large, or (ii) $G^{-1}$ has an anomalously small projection onto $Q | \sigma \rangle$, which is impossible since $G$ is nominally independent of geometry at the scale of the contacts, in particular the scale $\kappa$. In fact neither of these possibilities occurs: the mechanism by which $|f \rangle \sim p$, as required, is that the vector $Q | \sigma \rangle$ must be $\OO(\Delta)$ \textit{everywhere in the packing}. This is equivalent to the statement that all grain radii and grain positions must be specified to a precision $\OO(\Delta)$, the scale of particle deflections. If one had access to such data, one could determine the normal forces directly, without invoking Eq.\eqref{1}. 

\begin{figure}[th!]
\centering
\includegraphics[width=\columnwidth]{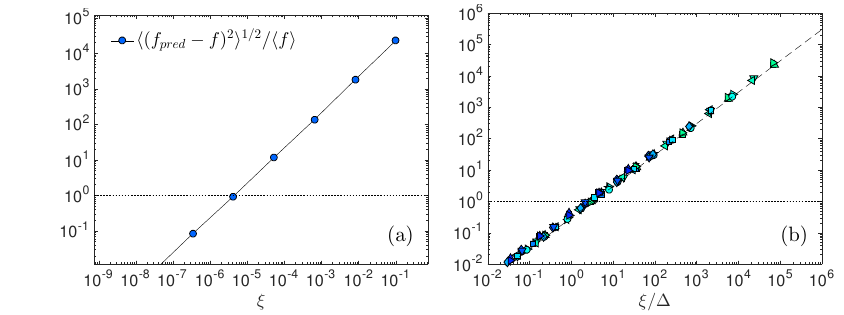}
\caption{ (a) Relative error in predicted forces versus noise $\xi$, for $\Delta \approx 10^{-6}$. (b) Collapse of all data $\langle (f_{pred}-f)^2\rangle^{1/2}/\langle f \rangle$ versus $\xi/\Delta$. Dashed line is $0.3 \; \xi/\Delta$.}\label{fig1a}
\end{figure}

To quantitatively demonstrate our result, we implemented Eq.\eqref{1} and tested its susceptibility to simulated experimental noise. With a standard DEM code we prepared packings over a range of $\Delta$ from $10^{-6}$ to $10^{-1}$ \footnote{For more details on our numerical procedure, see 
\cite{DeGiuli15b}. 
Rather than shearing our initial packings, as considered in Ref.\cite{DeGiuli15b}, here we added viscous drag and let the packings relax under constant external forces until the mean particle velocity \unexpanded{$\langle v\rangle< 10^{-8} \langle \sigma \rangle \sqrt{P/\rho}$}.}. 
We then added Gaussian noise of amplitude $\xi$ to the grain radii, and measured the relative error in predicted contact forces $f_{pred}$ from the true ones, $f$ (Fig 1). As expected from Eq.\eqref{1}, only when $\xi \lesssim \Delta$ is the relative error $\langle (f_{pred}-f)^2 \rangle^{1/2}/f$ much smaller than 1; for larger noise the predicted forces are not even of the correct magnitude.

We have shown that Eq.\eqref{1} is not useful for obtaining forces from geometrical information. If, somehow, the normal forces are known, then the mechanical equilibrium equations can be used to obtain the transverse forces, showing that the transverse force law is indeed redundant. This is relevant to the practical, unresolved problem: to determine which \textit{statistical} information about the packing is necessary to determine the \textit{macroscopic} stress response.



\bibliography{../bib/Wyartbibnew}

\begin{thebibliography}{9}%
\makeatletter
\providecommand \@ifxundefined [1]{%
 \@ifx{#1\undefined}
}%
\providecommand \@ifnum [1]{%
 \ifnum #1\expandafter \@firstoftwo
 \else \expandafter \@secondoftwo
 \fi
}%
\providecommand \@ifx [1]{%
 \ifx #1\expandafter \@firstoftwo
 \else \expandafter \@secondoftwo
 \fi
}%
\providecommand \natexlab [1]{#1}%
\providecommand \enquote  [1]{``#1''}%
\providecommand \bibnamefont  [1]{#1}%
\providecommand \bibfnamefont [1]{#1}%
\providecommand \citenamefont [1]{#1}%
\providecommand \href@noop [0]{\@secondoftwo}%
\providecommand \href [0]{\begingroup \@sanitize@url \@href}%
\providecommand \@href[1]{\@@startlink{#1}\@@href}%
\providecommand \@@href[1]{\endgroup#1\@@endlink}%
\providecommand \@sanitize@url [0]{\catcode `\\12\catcode `\$12\catcode
  `\&12\catcode `\#12\catcode `\^12\catcode `\_12\catcode `\%12\relax}%
\providecommand \@@startlink[1]{}%
\providecommand \@@endlink[0]{}%
\providecommand \url  [0]{\begingroup\@sanitize@url \@url }%
\providecommand \@url [1]{\endgroup\@href {#1}{\urlprefix }}%
\providecommand \urlprefix  [0]{URL }%
\providecommand \Eprint [0]{\href }%
\providecommand \doibase [0]{http://dx.doi.org/}%
\providecommand \selectlanguage [0]{\@gobble}%
\providecommand \bibinfo  [0]{\@secondoftwo}%
\providecommand \bibfield  [0]{\@secondoftwo}%
\providecommand \translation [1]{[#1]}%
\providecommand \BibitemOpen [0]{}%
\providecommand \bibitemStop [0]{}%
\providecommand \bibitemNoStop [0]{.\EOS\space}%
\providecommand \EOS [0]{\spacefactor3000\relax}%
\providecommand \BibitemShut  [1]{\csname bibitem#1\endcsname}%
\let\auto@bib@innerbib\@empty
\bibitem [{\citenamefont {Agnolin}\ and\ \citenamefont
  {Roux}(2007)}]{Agnolin07a}%
  \BibitemOpen
  \bibfield  {author} {\bibinfo {author} {\bibfnamefont {I.}~\bibnamefont
  {Agnolin}}\ and\ \bibinfo {author} {\bibfnamefont {J.-N.}\ \bibnamefont
  {Roux}},\ }\href {\doibase 10.1103/PhysRevE.76.061302} {\bibfield  {journal}
  {\bibinfo  {journal} {Phys. Rev. E}\ }\textbf {\bibinfo {volume} {76}},\
  \bibinfo {pages} {061302} (\bibinfo {year} {2007})}\BibitemShut {NoStop}%
\bibitem [{\citenamefont {Atman}\ \emph {et~al.}(2013)\citenamefont {Atman},
  \citenamefont {Claudin}, \citenamefont {Combe},\ and\ \citenamefont
  {Mari}}]{Atman13}%
  \BibitemOpen
  \bibfield  {author} {\bibinfo {author} {\bibfnamefont {A.~P.}\ \bibnamefont
  {Atman}}, \bibinfo {author} {\bibfnamefont {P.}~\bibnamefont {Claudin}},
  \bibinfo {author} {\bibfnamefont {G.}~\bibnamefont {Combe}}, \ and\ \bibinfo
  {author} {\bibfnamefont {R.}~\bibnamefont {Mari}},\ }\href@noop {} {\bibfield
   {journal} {\bibinfo  {journal} {EPL (Europhysics Letters)}\ }\textbf
  {\bibinfo {volume} {101}},\ \bibinfo {pages} {44006} (\bibinfo {year}
  {2013})}\BibitemShut {NoStop}%
\bibitem [{\citenamefont {Edwards}\ and\ \citenamefont
  {Grinev}(2001)}]{Edwards01}%
  \BibitemOpen
  \bibfield  {author} {\bibinfo {author} {\bibfnamefont {S.~F.}\ \bibnamefont
  {Edwards}}\ and\ \bibinfo {author} {\bibfnamefont {D.~V.}\ \bibnamefont
  {Grinev}},\ }\bibfield  {booktitle} {\emph {\bibinfo {booktitle} {Proc. Int.
  Workshop on Frontiers in the Physics of Complex Systems}},\ }\href@noop {}
  {\bibfield  {journal} {\bibinfo  {journal} {Physica A: Statistical Mechanics
  and its Applications}\ }\textbf {\bibinfo {volume} {302}},\ \bibinfo {pages}
  {162} (\bibinfo {year} {2001})}\BibitemShut {NoStop}%
\bibitem [{\citenamefont {Ball}\ and\ \citenamefont
  {Blumenfeld}(2002)}]{Ball02}%
  \BibitemOpen
  \bibfield  {author} {\bibinfo {author} {\bibfnamefont {R.~C.}\ \bibnamefont
  {Ball}}\ and\ \bibinfo {author} {\bibfnamefont {R.}~\bibnamefont
  {Blumenfeld}},\ }\href@noop {} {\bibfield  {journal} {\bibinfo  {journal}
  {Phys. Rev. Lett.}\ }\textbf {\bibinfo {volume} {88}},\ \bibinfo {pages}
  {115505} (\bibinfo {year} {2002})}\BibitemShut {NoStop}%
\bibitem [{\citenamefont {DeGiuli}\ and\ \citenamefont
  {Schoof}(2014)}]{DeGiuli14a}%
  \BibitemOpen
  \bibfield  {author} {\bibinfo {author} {\bibfnamefont {E.}~\bibnamefont
  {DeGiuli}}\ and\ \bibinfo {author} {\bibfnamefont {C.}~\bibnamefont
  {Schoof}},\ }\href {http://stacks.iop.org/0295-5075/105/i=2/a=28001}
  {\bibfield  {journal} {\bibinfo  {journal} {EPL (Europhysics Letters)}\
  }\textbf {\bibinfo {volume} {105}},\ \bibinfo {pages} {28001} (\bibinfo
  {year} {2014})}\BibitemShut {NoStop}%
\bibitem [{\citenamefont {Gendelman}\ \emph {et~al.}(2016)\citenamefont
  {Gendelman}, \citenamefont {Pollack}, \citenamefont {Procaccia},
  \citenamefont {Sengupta},\ and\ \citenamefont {Zylberg}}]{Gendelman16}%
  \BibitemOpen
  \bibfield  {author} {\bibinfo {author} {\bibfnamefont {O.}~\bibnamefont
  {Gendelman}}, \bibinfo {author} {\bibfnamefont {Y.~G.}\ \bibnamefont
  {Pollack}}, \bibinfo {author} {\bibfnamefont {I.}~\bibnamefont {Procaccia}},
  \bibinfo {author} {\bibfnamefont {S.}~\bibnamefont {Sengupta}}, \ and\
  \bibinfo {author} {\bibfnamefont {J.}~\bibnamefont {Zylberg}},\ }\href
  {http://link.aps.org/doi/10.1103/PhysRevLett.116.078001} {\bibfield
  {journal} {\bibinfo  {journal} {Physical Review Letters}\ }\textbf {\bibinfo
  {volume} {116}},\ \bibinfo {pages} {078001} (\bibinfo {year}
  {2016})}\BibitemShut {NoStop}%
\bibitem [{Note1()}]{Note1}%
  \BibitemOpen
  \bibinfo {note} {For Hertzian contacts, $\Delta =(p/E)^{2/3}$, where $E$ is
  the grain Young modulus. Within a 20cm deep sandpile, $\Delta ~\approx
  ~10^{-5}$}\BibitemShut {NoStop}%
\bibitem [{Note2()}]{Note2}%
  \BibitemOpen
  \bibinfo {note} {For more details on our numerical procedure, see \cite
  {DeGiuli15b}. Rather than shearing our initial packings, as considered in
  Ref.\cite {DeGiuli15b}, here we added viscous drag and let the packings relax
  under constant external forces until the mean particle velocity $\langle
  v\rangle < 10^{-8} \langle \sigma \rangle \sqrt {P/\rho }$.}\BibitemShut
  {Stop}%
\bibitem [{\citenamefont {DeGiuli}\ \emph {et~al.}(2015)\citenamefont
  {DeGiuli}, \citenamefont {McElwaine},\ and\ \citenamefont
  {Wyart}}]{DeGiuli15b}%
  \BibitemOpen
  \bibfield  {author} {\bibinfo {author} {\bibfnamefont {E.}~\bibnamefont
  {DeGiuli}}, \bibinfo {author} {\bibfnamefont {J.}~\bibnamefont {McElwaine}},
  \ and\ \bibinfo {author} {\bibfnamefont {M.}~\bibnamefont {Wyart}},\
  }\href@noop {} {\bibfield  {journal} {\bibinfo  {journal} {arXiv preprint
  arXiv:1509.03512}\ } (\bibinfo {year} {2015})}\BibitemShut {NoStop}%
\end{thebibliography}%

\end{document}